\documentclass{jfm}

\usepackage{graphicx}
\usepackage{epsfig}
\usepackage{natbib}

\ifCUPmtlplainloaded \else
  \checkfont{eurm10}
  \iffontfound
    \IfFileExists{upmath.sty}
      {\typeout{^^JFound AMS Euler Roman fonts on the system,
                   using the 'upmath' package.^^J}%
       \usepackage{upmath}}
      {\typeout{^^JFound AMS Euler Roman fonts on the system, but you
                   dont seem to have the}%
       \typeout{'upmath' package installed. JFM.cls can take advantage
                 of these fonts,^^Jif you use 'upmath' package.^^J}%
      }
  \else
  \fi
\fi

\ifCUPmtlplainloaded \else
  \checkfont{msam10}
  \iffontfound
    \IfFileExists{amssymb.sty}
      {\typeout{^^JFound AMS Symbol fonts on the system, using the
                'amssymb' package.^^J}%
       \usepackage{amssymb}%
       \let\le=\leqslant  
         
      }{}
  \fi
\fi

\ifCUPmtlplainloaded \else
  \IfFileExists{amsbsy.sty}
    {\typeout{^^JFound the 'amsbsy' package on the system, using it.^^J}%
     \usepackage{amsbsy}}
    {\providecommand\boldsymbol[1]{\mbox{\boldmath $##1$}}}
\fi

\providecommand\bnabla{\boldsymbol{\nabla}}
\providecommand\bcdot{\boldsymbol{\cdot}}

\newsavebox{\astrutbox}
\sbox{\astrutbox}{\rule[-5pt]{0pt}{20pt}}

\newcommand{\mr}[1]{\mathrm{#1}}
\newcommand{\mb}[1]{\mathbf{#1}}
\def\<{\left\langle} \def\>{\right\rangle} \def\({\left(} \def\){\right)}
\def\[{\left[}
\def\]{\right]}
\newcommand{\bsv}{\boldsymbol{v}}
\newcommand{\bsF}{\boldsymbol{F}}
\newcommand{\bs}[1]{\boldsymbol{#1}}
\newcommand{\mri}{\mathrm{i}}
\newcommand{\KC}{Kistovich and Chashechkin }
\newcommand{\om}{\omega_{\mathrm{M2}}}

\title{Propagating and evanescent internal waves in a deep ocean model}

\author[M. S. Paoletti and Harry L. Swinney]%
{M.\ns S.\ns P\ls A\ls O\ls L\ls E\ls T\ls T\ls I \and
H\ls A\ls R\ls R\ls Y\ns L.\ns S\ls W\ls I\ls N\ls N\ls E\ls Y%
  \thanks{Email address for correspondence: swinney@chaos.utexas.edu}}

\affiliation{Department of Physics and Center for Nonlinear Dynamics, University of Texas at Austin, Austin, TX, 78712, USA}

\pubyear{2012}
\volume{??}
\pagerange{?--?}
\date{?; revised ?; accepted ?. - To be entered by editorial office}
\begin{document}

\maketitle

\begin{abstract}
We present experimental and computational studies of the propagation of internal waves in a stratified fluid with an exponential density profile that models the deep ocean.  The buoyancy frequency profile $N(z)$ (proportional to the square root of the density gradient) varies smoothly by more than an order of magnitude over the fluid depth, as is common in the deep ocean.  The nonuniform stratification is characterized by a turning depth $z_c$, where $N(z_c)$ is equal to the wave frequency $\omega$ and $N(z < z_c) < \omega$.  Internal waves reflect from the turning depth and become evanescent below the turning depth.  The energy flux below the turning depth is shown to decay exponentially with a decay constant given by $ k_c$, which is the horizontal wavenumber at the turning depth.  The viscous decay of the vertical velocity amplitude of the incoming and reflected waves above the turning depth  agree within a few percent with a previously untested theory for a fluid of arbitrary stratification [Kistovich and Chashechkin, J. App. Mech. Tech. Phys. \textbf{39}, 729-737 (1998)].   
\end{abstract}

\section{Introduction}\label{sec:Intro}
In a stratified fluid, an internal wave transports momentum and energy as the propagating disturbance is restored by buoyancy forces.  In the absence of rotation, internal waves satisfy the following dispersion relation
\begin{equation}
\label{eq:Dispersion_relation}
\frac{\omega}{N(z)} = \sin \(\theta (z)\),
\end{equation}
where $\omega$ is the angular frequency of the wave, $\theta$ is the angle of propagation relative to the horizontal, $N(z) = \sqrt{-\(g/\rho_0\)(d\rho/dz)}$ is the buoyancy frequency, $g$ is the gravitational acceleration, $\rho_0$ is a reference density and $\rho(z)$ is the density profile.  Approximately half of the internal wave energy in the ocean is produced by tidal flow over bottom topography \citep{munk98,wunsch04}. 

Oceanic buoyancy frequencies vary greatly, decreasing from the values in shallow water to become two orders of magnitude smaller in the well-mixed abyss (see example in \S\ref{sec:Results}).  \citet{pingree91} found that deep in the Bay of Biscay $N(z)$ became so small that the internal waves became nearly vertical (Eq.\ \ref{eq:Dispersion_relation}) at the point of bottom boundary reflection. Recently \citet{king12}  analyzed data for pressure, temperature, and salinity for thousands of locations throughout the oceans and found that in many locations in the deep oceans the buoyancy frequency becomes smaller than $\om = 1.4052 \times 10^{-4}$~rad/s, the frequency of the M$_2$ (lunar) tides. Internal waves approaching  a {\em turning depth} $z_c$ where $N(z_c) = \om$  become vertical by (\ref{eq:Dispersion_relation}) and then reflect from the turning surface; below the turning depth the waves are evanescent (exponentially damped).   

\citet{kistovich98} developed a linear theory that describes the propagation of internal waves in arbitrary stratifications, including those that contain a turning depth.  While their theory gives analytic expressions for the vertical velocity field, the derivation requires that the wavelength and beam width be small compared to the length scale characterizing the stratification, which is the case examined here.  Recent studies \citep{nault07,mathur09,mathur10} have developed theories for the propagation of internal waves in nonuniform stratifications that do not require these assumptions.  \citet{sutherland04,gregory10} performed experimental tests with a focus on the reflection and transmission of internal waves through weakly stratified regions.  \citet{mathur09,mathur10} verified their theory with experiments focused on the interaction of internal wave beams with sharp density gradients and finite-width transition regions, as a model of the upper-ocean and thermocline.   

Here, we focus on internal wave propagation in stratified fluids with exponentially varying buoyancy frequency profiles $N(z)$ to model the deep ocean (see example in \S\ref{sec:Results}).  While the buoyancy frequencies in our studies vary by a factor of 12, the variation is sufficiently slow that the assumptions of \citet{kistovich98} are satisfied, allowing us to perform the first tests of their theory.  Our experiments and numerical simulations examine reflection from a turning depth and the viscous decay of internal waves as they propagate in nonuniform stratifications.  In addition, we characterize the decay of the energy and energy flux for evanescent waves beneath a turning depth.    

\section{Kistovich-Chashechkin theory}\label{sec:Prev_Theory}
\citet{kistovich98} describe the propagation of internal waves in arbitrary stratifications.  The two-dimensional (2D) theory accounts for viscous dissipation, diffusion and turning depths.  The density profile is assumed to be $\rho_0(z) = \rho_{00}\[1 + s_0 (z)\]$, where $s_0$ is the reduced salinity and $\rho_{00}$ is a reference density.  The internal wavefield is assumed to have a time-dependence of the form $\exp (-\mri \omega t)$ (which is henceforth omitted). Then the linear Navier-Stokes equations in the Boussinesq approximation become
\begin{eqnarray}
-\mri \omega u = -\frac{1}{\rho_{00}}\frac{\partial p}{\partial x} + \nu \Delta u,&\quad&
-\mri \omega w = -\frac{1}{\rho_{00}}\frac{\partial p}{\partial z} + \nu \Delta w - sg\\
-\mri \omega s + w\frac{ds_0}{dz} = D \Delta s,&\quad&
\frac{\partial u}{\partial x} + \frac{\partial w}{\partial z} = 0,
\end{eqnarray}
where $u$ is the horizontal velocity, $w$ the vertical velocity, $p$ and $s$ the pressure and salinity, $\nu$ the kinematic viscosity, $D$ the salt diffusivity, $g$ gravitational acceleration (in the direction $-\hat{z}$), and $\Delta = \partial ^2/\partial x^2 + \partial ^2/\partial z^2$ is the two-dimensional Laplacian.

The internal wavefield may be described by the vertical displacement $h(x,z)$, which is related to the vertical velocity by $w = -\mri \omega h$.  Solutions may be sought in the form
\begin{equation}
h(x,z) = \int_0^{\infty} f(z,k)\exp (\mri kx) dk,
\end{equation}
where $k$ is the wavenumber.

Asymptotic solutions for $f$ may be obtained if $N(z)$ varies slowly compared to the length scales describing the internal wave beam, yielding
\begin{equation}
f(z,k) = \frac{A(k)}{\sqrt{\gamma}} \exp \[ \frac{\mri \tilde{\nu} k^2 \(1 - 3\gamma^4\)}{4 \omega \gamma ^2}\]\exp\left\{\mri \xi k \int_{z_0}^z \[ \gamma - \frac{\mri \tilde{\nu} k^2 \(1 + \gamma^2 \)^2}{2 \omega \gamma}\] dz^{\prime} \right\},
\label{eq:f_i}
\end{equation}
where $A(k)$ describes the spectral properties of the wave source located at $\(x_0, z_0\)$, $\tilde{\nu} = \nu + D$, $\xi = 1$ ($-1$) for beams propagating down (up), and $\gamma ^2(z) = (N^2(z) - \omega^2)/\omega^2$.  

 In addition, \KC describe the reflection of an internal wave beam from a turning depth $z_c$, defined as $N(z_c) = \omega$ and $N(z) < \omega$ for all $z < z_c$:  
 \begin{eqnarray} \label{eq:f_r} f_r (z,k) = \frac{B(k)}{\sqrt{|\gamma|}}\exp \[ \frac{\mri \tilde{\nu} k^2\(1- 3\gamma^4\)}{4 \omega \gamma^2} - \mri k\int_{z_c}^z\left\{\gamma - \frac{\mri \tilde{\nu} k^2 \(1 + \gamma^2\)^2}{2\omega \gamma}\right\}dz^{\prime}\],\\
\label{eq:f_t} f_t (z,k) = \frac{C(k)}{\sqrt{|\gamma|}}\exp \[ \frac{\mri \tilde{\nu} k^2\(1- 3\gamma^4\)}{4 \omega \gamma^2} +  k\int_{z_c}^z\left\{|\gamma| + \frac{\mri \tilde{\nu} k^2 \(1 + \gamma^2\)^2}{2\omega |\gamma|}\right\}dz^{\prime}\],
 \end{eqnarray}
 where $f_r$ is the reflected wave for $x > x_c$ and $z > z_c$ and $f_t$ is the transmitted (evanescent) wave for $z < z_c$.  Given the spectral amplitudes $A(k)$ of the incoming wave, we have $B(k) = A(k) e^{-\mri \phi_0} e^{\mri \pi/2}$, and $C(k) = A(k) e^{-\mri \phi_0} e^{\mri \pi/4}$, where
 \begin{equation}
\phi_0 = k\int_{z_0}^{z_c} \[ \gamma - \frac{\mri \tilde{\nu}k^2\(1+\gamma^2\)^2}{2\omega \gamma}\]dz^{\prime},
\label{eq:phi_0}
 \end{equation}
 which takes into account the phase winding and viscous decay as the beam propagates from the source at $\(x_0,z_0\)$ to the turning depth at $\(x_c,z_c\)$.  
\section{Methods}\label{sec:Methods}
\subsection{Two-dimensional numerical simulations of the Navier-Stokes equations}
We numerically simulate internal wave propagation in nonuniform stratifications by solving the Navier-Stokes equation using CDP~2.4 \citep{ham04}, which is a parallel, unstructured, finite-volume based solver modeled after  the algorithm of \citet{mahesh04}.   We disable the subgrid-scale modeling in CDP.  Second-order accuracy in space and time is achieved by using a fractional step time marching scheme and multiple implicit schemes for the spatial operators \citep{ham06}.  We perform 2D simulations in the Boussinesq approximation, which is justified by the small density variation and predominantly 2D flows examined in the experiments.  CDP solves the following equations for the pressure $p$, density $\rho$ and velocity $\bsv = (u,w)$ in the $(x,z)$ directions:
\begin{eqnarray}
\label{eq:NS}
\partial{\bsv}/\partial{t} + (\bsv \bcdot \bnabla)\bsv = -\nabla p/\rho_0 - g\rho/\rho_0\mb{\hat{z}} + \nu \bnabla^2 \bsv + \bs{F}/\rho_0,\\
\label{eq:incomp} \bnabla \bcdot \bsv = 0,\quad
\label{eq:dens_advec_diff} \partial{\rho}/\partial{t} + \(\bsv \bcdot \bnabla \)\rho = D \nabla^2 \rho,
\end{eqnarray}
where $\rho_0 = 1000$~kg/m$^3$ is a reference density, $g$ the gravitational acceleration, $\nu$ the kinematic viscosity, and $\bsF$ drives the internal waves, as described below.  The density diffusion coefficient $D = 2 \times 10^{-9}$~m$^2$/s is equal to the value for sodium chloride, which is used in the experiments, resulting in a Schmidt number $\nu/D = 500$ in water.  Simulations for Schmidt number values varying from 10 to 10$^6$ yielded the same results within 1.5\%, as expected for our low Reynolds number system where there is no wave breaking.

Internal waves are generated by the forcing term $\bsF$ in Eq.~(\ref{eq:NS}), which is similar to that used by \citet{slinn98}, \citet{javam99} and \citet{rodenborn11}:
\begin{eqnarray}
\bsF (x,z) =\( \bnabla \times \bnabla \times \Psi \hat{z} \) \sin (\omega t),\\ \label{eq:F}
\Psi (x,z) = \Psi_0 \exp \[-\(a \(x - x_0\)^2 + 2b\(x-x_0\)\(z-z_0\) + c\(z-z_0\)^2\)\], \label{eq:Psi}
\end{eqnarray}
where $a = \cos^2(\beta)/2\sigma_x^2 + \sin^2(\beta)/2\sigma_z^2$, $b = \sin(2\beta)/4\sigma_x^2 - \sin(2\beta)/4\sigma_z^2$, $c = \sin^2(\beta)/2\sigma_x^2 + \cos^2(\beta)/2\sigma_z^2$, $\omega$ is the internal wave frequency, $\Psi_0$ the maximum forcing amplitude, $\(x_0, z_0\)$ the spatial center, $\beta$ the counterclockwise rotation of the elliptical Gaussian forcing profile, and $\sigma_x$ ($\sigma_z$) is the horizontal (vertical) standard deviation.  The angle $\beta$ is chosen so that the forcing profile is aligned with the local internal wave propagation angle $\theta$.  The oscillation frequency is $\omega_0 = 0.628$~rad/s for all of the simulations and experiments except for Run~12 where $\omega = \omega_0/4$ (see table \ref{table:Params} in online supplementary material).  The time step is $\Delta t = \pi/4000 \omega$, which yields 8000 time steps per period.  Such fine time resolution is required to reduce the numerical dissipation enough that accurate measurements of the viscous decay are possible.  We run the simulations for 40 periods to ensure that the system has reached a steady-state.  The forcing profile $\Psi(x,z)$ is varied to modify the properties of the wave beam (see table \ref{table:Params} in online supplementary material).  

Two computational grids are generated using Pointwise Gridgen.  Grid I is designed for direct comparison with the experiments.  The domain has the same dimensions as the experimental tank, namely $0 < x < 0.9$~m and $0 < z < 0.6$~m.  The wavemaker is centered at $(x_0 = 0.15$~m, $z_0 = 0.50$~m$)$ in both the simulations and experiments.  While the grid is structured (composed of rectangular control volumes), the horizontal and vertical resolutions vary smoothly to take into account the variable shear as the beam propagates in the nonlinear stratification.  Near the wavemaker the wave beam is nearly horizontal, resulting in strong vertical shear, while at the turning depth the beam propagates vertically, necessitating increased horizontal resolution.  At the wavemaker we use a horizontal resolution of $\Delta x = 2.5$~mm, while the vertical resolution is $\Delta z = 0.4$~mm.  The resolutions near the reflection from the turning depth at $\(x_c, z_c\)$ are $\Delta x = 0.4$~mm and $\Delta z = 1.5$~mm.  Grid~I is composed of approximately $5.3 \times 10^5$ control volumes.  No-slip boundary conditions are enforced at $x = 0,$ $z = 0$, and $x = 0.90$~m.  The top surface at $z = 0.60$~m is a free-slip, impenetrable boundary chosen to emulate the free surface in the experiments.  The internal waves are damped for $x > 0.8$~m by adding a drag force equal to $-0.002 \bsv$ to Eq.(\ \ref{eq:NS}) to mimic the absorbent material present in the experiments.

Grid~II is much larger, which allows for greater viscous decay.  This increased propagation distance provides a better testbed for the theoretical predictions of \citet{kistovich98}, as any discrepancies will be magnified.  The grid spans $0 < x < 6$~m and $0 < z < 2$~m.  The wavemaker is centered at $(x_0 = 1.0$~m, $z_0 = 1.5$~m$)$.  As with Grid~I, the spatial resolution varies smoothly throughout the domain to account for the variation in the beam angle as it propagates.  The horizontal resolution in the vicinity of the wavemaker is $\Delta x = 5$~mm while the vertical resolution is $\Delta z = 0.625$~mm.  Near the reflection at the turning depth the horizontal resolution is $\Delta x = 1$~mm and the vertical resolution is $\Delta z = 2.5$~mm.  Grid~II has about $2.5 \times 10^6$ control volumes.  No-slip boundary conditions are applied at all of the boundaries.  To reduce finite-size effects, a drag force of the form $-0.02 \bsv$ is added to Eq.\ \ref{eq:NS} for $x < 0.6$~m, $x > 5.5$~m and $z > 1.75$~m to damp the wave beams before they reach the boundaries.  We observe less than a 1\% change in our computed velocity fields upon doubling and quadrupling the temporal and spatial resolution in our simulations, thereby exhibiting convergence of our solutions.   

The density profiles are chosen to be scaled models of the stratification in the deep ocean, which are typically exponential (see example in \S\ref{sec:Results}), that have a turning depth at $z_c$ with $N(z) < \omega$ for $z < z_c$.  However, our results do not depend upon the stratification being exponential in form; the crucial aspect is the existence of a turning depth.  Two stratifications are used for the experimental domain covered by Grid~I, namely
\begin{eqnarray}
\label{eq:rho_I}
\rho_{\mr{expt1}} = 1061 - 10\exp(3.12z) \mathrm{~kg/m}^3, \quad 0 \le z \le 0.6 \mr{~m}, \\ 
\rho_{\mr{expt2}} = 1054 - 0.189\exp(11.4z) \mr{~kg/m}^3, \quad 0 \le z \le 0.5 \mr{~m}, \label{eq:rho_expt2}
\end{eqnarray}
The depth-dependent buoyancy frequency $N_{\mr{expt1}}(z)$ has a turning depth at $z_c = 0.12$~m, while the turning depth for $N_{\mr{expt2}}(z)$ is at $z_c = 0.257$~m.  The density profile $\rho_{\mr{expt1}}$ is used primarily to study the incoming and reflected wave beams, while the higher turning depth for $\rho_{\mr{expt2}}$ allows for more detailed characterization of the evanescent waves.  The density profile used for all simulations with Grid~II, except Run~12, is given by
\begin{equation}
  \rho_{\mr{sim}} = 1244-5.75\exp (2.5z), \quad 0 \le z \le 2 \mr{~m}
\label{eq:rho_II}
\end{equation}
The resulting buoyancy frequency profile varies by a factor of 12 from top to bottom and has a turning depth at $z_c = 0.424$~m.  The density profile for Run~12 has $N(z)$ reduced by a factor of 4 for all depths. 
\subsection{Experimental Methods}\label{subsec:Experimental_Methods}
Experiments are performed in a glass tank with $0 < x < 0.90$~m, $0 < y < 0.45$~m and $0<z<0.60$~m.  The exponential gradients given by Eqs.\ (\ref{eq:rho_I}) and (\ref{eq:rho_expt2}) are generated by utilizing the generalized version of the double-bucket method described by \citet{hill02}.  The density profile is measured by extracting fluid samples at known heights, which are then measured with an Anton Paar density meter.  Example experimental stratification measurements are given in tables \ref{table:rho_expt1} and \ref{table:rho_expt2}.  The experimental density profiles agree well with the stratification used in the simulations (Eqs.\ (\ref{eq:rho_I}) and(\ref{eq:rho_expt2})) for $0.1 < z < 0.5$~m.

Internal wave beams are generated using a wave maker similar to the one developed by \citet{gostiaux07}.  Our wave maker \citep{rodenborn11} consists of five acrylic plates with dimensions 150~mm~$\times$~150~mm~$\times$~6~mm that are separated by $2.5$~mm and placed within a parallelpiped open-sided box.  A helicoidal rotating camshaft oscillates the plates periodically.  The eccentricity of the camshaft ($12.7$~mm) determines the oscillation amplitude.  The generated wave beams are nearly two-dimensional, particularly along the center of the tank ($y = 0.225$~m) where all of our measurements are taken.  The Reynolds number of the generated waves is $Re \approx 150$, where we have used the wavelength and maximum speed as the relevant length and velocity scales.  The local Richardson number varies along the beam, but has a minimum value of $Ri \approx 100$.  The wave maker is centered at $(x = 0.15$~m, $z = 0.50$~m$)$ and oriented parallel to the the local wave beam angle.  The internal wave motions are dissipated by a fiber mesh for $x > 0.88$~m.

Particle image velocimetry is used to obtain the two-dimensional velocity field $\bsv = (u,w)$ in the vertical plane given by $y = 0.225$~m.  Hollow glass spheres with diameters $8 < d < 12$~$\mu$m and densities in the range $1050 < \rho < 1150$~kg/m$^3$ are used as seed particles. The tracers are illuminated by a 5~mm thick laser sheet produced by a 532~nm wavelength laser (2.5~W).  Two 12-bit CCD cameras with $1296 \times 966$ pixel resolution capture orthogonally scattered light in a $194 \times 145$~mm$^2$ region, 40 times per period $(\Delta t = 0.25$~s$)$.  The instantaneous velocity fields are determined using the CIV algorithm developed by \citet{fincham00} and are interpolated to a regular $50 \times 50$ grid with spatial resolutions of $\Delta x = 3.9$~mm and $\Delta z = 2.9$~mm.

\begin{figure}
  \centerline{\includegraphics[width=\columnwidth]{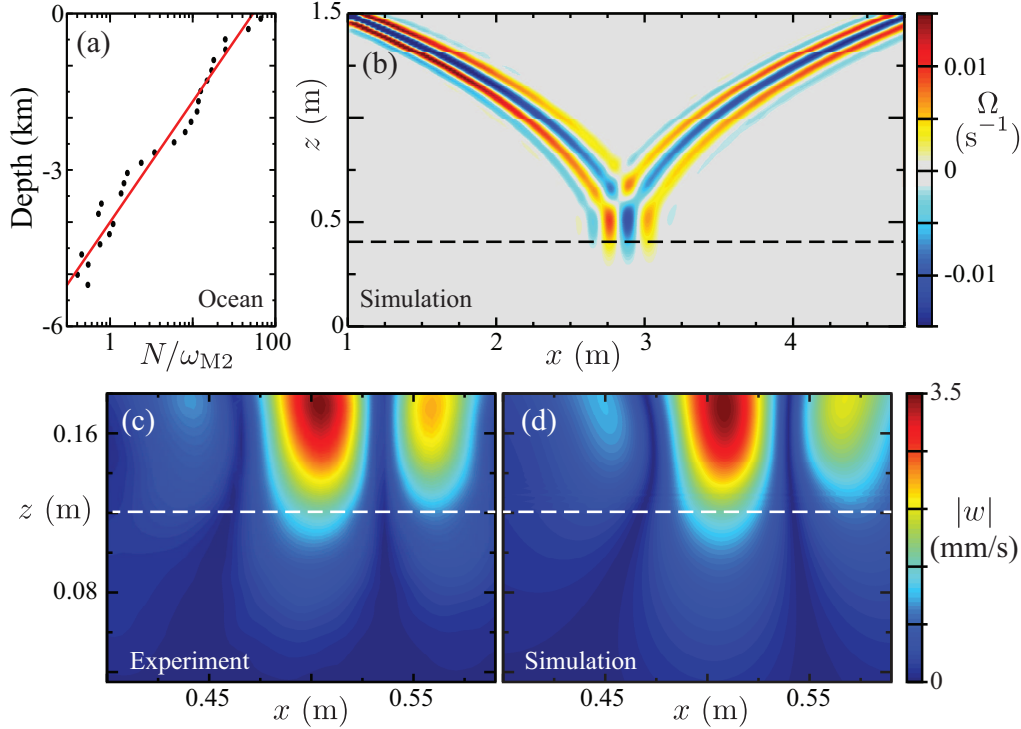}}
  \caption{(a) Buoyancy frequency profile computed using the method described by \citet{king12} at 13.02$^{\circ}$~N latitude and 91.77$^{\circ}$~W longitude (Transect P19C, World Ocean Circulation Experiment, February 1993).  The red line is an exponential fit to the data.  (b) Snapshot of the vorticity (color) for a wave beam generated at $(x = 1$~m, $z = 1.5$~m$)$ in simulations, after 40~periods of oscillation.  The stratification (Eq.\ (\ref{eq:rho_II})) has a turning depth at $z_c = 0.424$~m (dashed line).  Absolute value of the vertical velocity (color) near the turning depth (dashed line) in (c) experiments and (d) simulations with stratification (\ref{eq:rho_I}).}
\label{fig:vorticity}
\end{figure}

\section{Results} \label{sec:Results}
We have determined $N(z)$ for all 18100 data sets measured in the World Ocean Circulation Experiment (WOCE).  This analysis revealed that $N(z)$ commonly shows an approximate exponential variation in the deep oceans.  In the example shown in Fig.\ \ref{fig:vorticity}(a), $N(z)$ varies by two orders of magnitude, and a turning depth exists for $z_c \approx -4$~km. 

An example snapshot of a wave beam propagating in the stratification given by Eq.\ (\ref{eq:rho_II}) is shown in Fig.\ \ref{fig:vorticity}(b) (see movie in online supplementary material), with vorticity $\Omega = \bnabla \times \bsv$ indicated by color.  The beam generated at $(x = 1.0 $~m, $z = 1.5$~m$)$ propagates down and to the right before being reflected at the turning depth, causing the energy to move up and to the right.  Only a small fraction of the energy flux penetrates below the turning depth, as shown by snapshots of the vertical velocity in Figs.\ \ref{fig:vorticity}(b) and (c).

\begin{figure}
  \centerline{\includegraphics[width=\columnwidth]{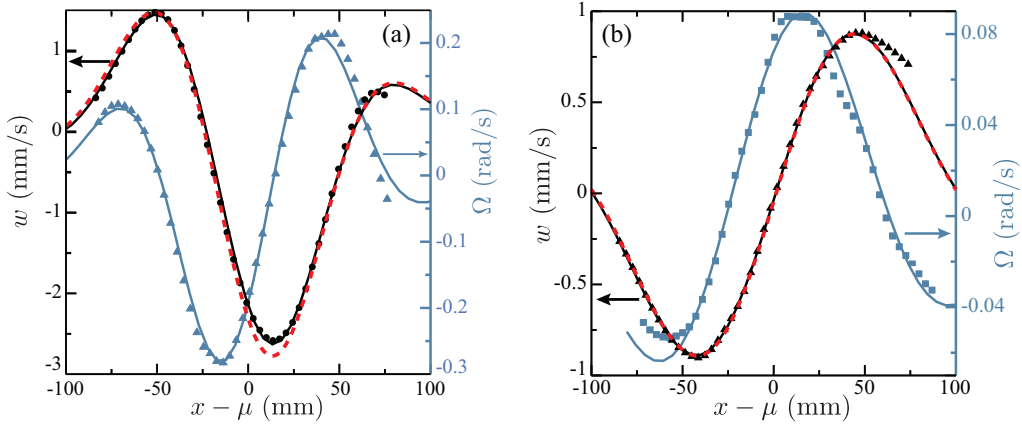}}
  \caption{Snapshots of the vertical velocity (black) and vorticity (blue) horizontal cross-sections for the (a) incoming wave beam centered at $(\mu =0.30$~m, $z =0.40$~m$)$ and (b) reflected wave beam at $(\mu = 0.76$~m, $z = 0.37$~m$)$ for the stratification given by Eq.\ (\ref{eq:rho_I}).  Laboratory measurements are given by the data points, results from the simulations are shown as solid black and blue lines, and theoretical predictions are denoted by the dashed red lines.}
\label{fig:vz_Xsecs}
\end{figure}

We obtain remarkable agreement between the experiments, simulations and theory, as illustrated by comparing Figs.\ \ref{fig:vorticity}(c) and (d) as well as the beam profiles in Fig.\ \ref{fig:vz_Xsecs}.  Horizontal profiles of $w$ are well described by modulated Gaussians of the form
\begin{equation}
w(x,z,t) = A_0(z) \[\gamma(z_0)/\gamma(z)\]^{1/2} \cos \(k_0(z)x - \omega t\) \exp \(-\[x-\mu(z)\]^2/2\sigma(z)^2\),
\label{eq:w_profile}
\end{equation}
where $A_0$ is the velocity amplitude, $k_0$ the wavenumber, $\mu$ the beam center, $\sigma$ the beam width, and the term $\[\gamma(z_0)/\gamma(z)\]^{1/2}$ accounts for the increase (decrease) in the vertical velocity as the beam propagates into weaker (stronger) stratification.  Even though the vertical velocity may increase by geometrical effects as $N(z)$ decreases, the velocity amplitude $A_0$ monotonically decreases owing to viscous dissipation.  

\subsection{Evanescent internal waves} \label{subsec:Evanescent}
\begin{figure}
  \centerline{\includegraphics[width=\columnwidth]{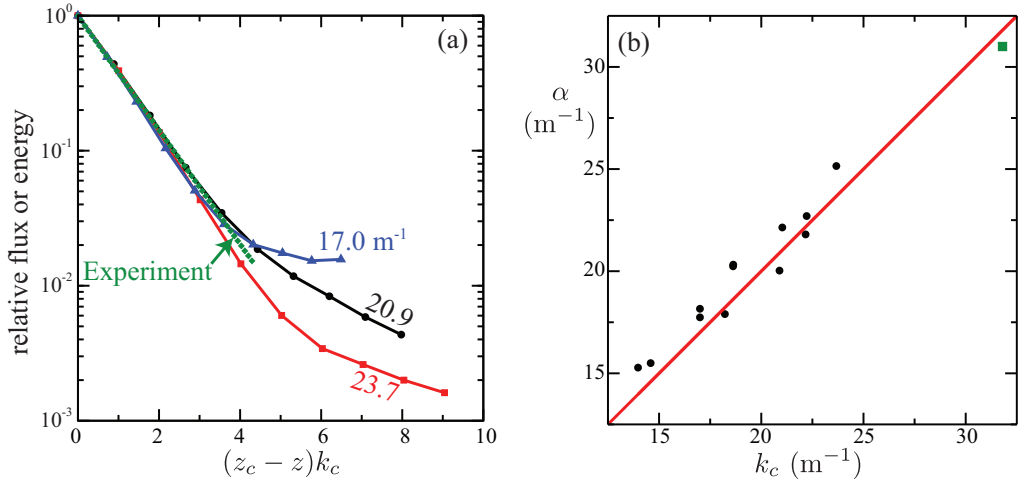}}
  \caption{(a) Decay of the relative horizontally integrated flux $\Phi(z)/\Phi(z_c)$ for evanescent waves in the simulations for the stratification given by Eq.\ (\ref{eq:rho_II}).  The horizontal wavenumber at the turning depth, $k_c$, is indicated next to each (black, red, and blue) curve.  Measurements of the relative horizontally integrated kinetic energy $E(z)/E(z_c)$ for the stratification (\ref{eq:rho_expt2}) are shown as green diamonds.  (b) The decay constant $\alpha$ from simulations (black circles) and the experiments (green square) are compared to the horizontal wavenumber at the turning depth, $k_c$ (red line).} 
\label{fig:flux_plots}
\end{figure}

Below the turning depth $z_c$, internal waves are evanescent with an intensity that decays exponentially with a decay constant $\alpha$ \citep{gostiaux06}.  We  compute the intensity of the evanescent wave beams by horizontally integrating the vertical energy flux, 
\begin{equation}
\Phi(z) = \int p^{\prime} w^{\prime}dx,
\label{eq:Phi}
\end{equation}
where $p^{\prime}$ and $w^{\prime}$ are the fluctuations in the pressure and vertical velocity, respectively.  Figure \ref{fig:flux_plots}(a) shows example results for the vertical energy flux below the turning depth from our simulations.  The distance from the turning depth $z_c - z$ is scaled by the horizontal wavenumber at the turning depth $k_c \equiv k_0(z_c)$.  Below the turning depth, the energy flux decays exponentially, except near the bottom-boundary.  We characterize the experimental evanescent waves by the horizontally integrated kinetic energy, $E(z) = 0.5 \int |\bsv|^2 dx$, which is also shown to decay exponentially in Fig. \ref{fig:flux_plots}(a) for the stratification (\ref{eq:rho_expt2}).

Our measurements of the decay constants $\alpha$ from both the simulations and experiments agree well with the expected values $\alpha = k_c$ (see Fig.\ \ref{fig:flux_plots}(b) and table \ref{table:Results}).  Therefore, in analogy with evanescence in other contexts, we conclude that the energy flux is given by 
\begin{equation}
\Phi(z) = \Phi(z_c)e^{-k_c (z_c-z)}, \quad \mathrm{for~} z < z_c.
\label{eq:Phi_vs_z}
\end{equation}
\subsection{Viscous decay and internal wave reflection: comparison with theory} \label{subsec:Theory_Compare}
The theory of \citet{kistovich98} (see \S\ref{sec:Prev_Theory}) provides predictions  with {\em no free parameters} for the vertical velocity field of internal waves propagating in arbitrary stratifications.  To test these predictions, we compare horizontal cross-sections of the vertical velocity field computed by our numerical simulations to the theoretical predictions.  Beam profiles are characterizing by fitting them to the form (\ref{eq:w_profile}) for varying $z$.

To use the theoretical expressions (\ref{eq:f_i}) and (\ref{eq:f_r}) for the incoming and reflected wave beams, we must know the buoyancy frequency profile $N(z)$ and the spectral amplitudes of the wave source $A(k)$.  The buoyancy frequency profile may be determined using Eq.\ (\ref{eq:rho_II}).  We determine $A(k)$ by fitting a horizontal cross-section of the vertical velocity near the wave source to Eq.\ (\ref{eq:w_profile}), which we then Fourier transform analytically to determine the spectral amplitudes.  We then compute the predicted vertical velocity field for the incoming and reflected wave beams using Eqs.\ (\ref{eq:f_i}) and (\ref{eq:f_r}).

\begin{figure}
  \centerline{\includegraphics[width=\columnwidth]{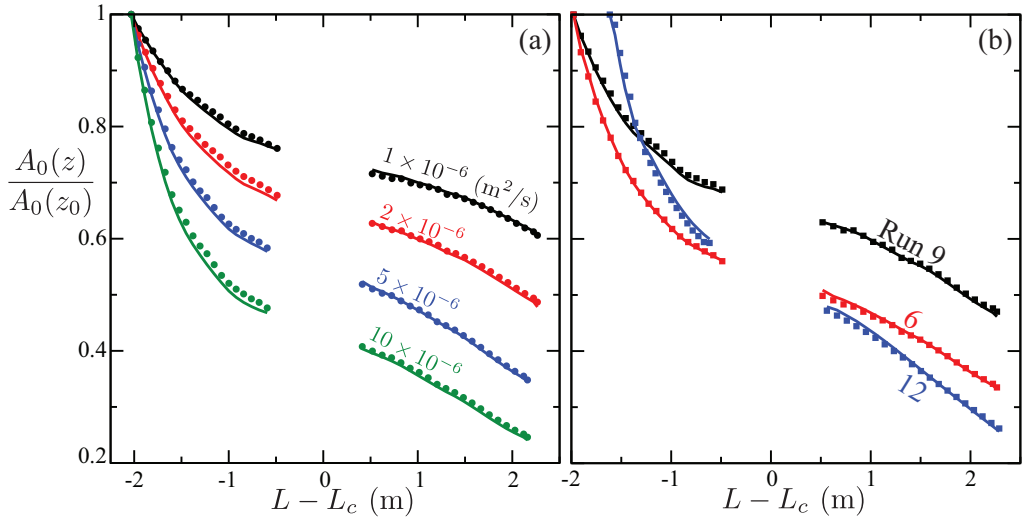}}
  \caption{The vertical velocity amplitude $A_0(z)$ in the simulations (points) is compared with the fully constrained theoretical predictions of \citet{kistovich98} (solid lines) for (a) varying viscosity values and (b) fixed viscosity ($5\times 10^{-6}$m$^2$/s) and other parameters given by ($\omega$, $k_0(z_0)$, $\sigma(z_0)$): Run 6, (0.63 rad/s, 22.3 m$^{-1}$, 0.127 m); Run 9, (0.63 rad/s, 18.7 m$^{-1}$, 0.144 m); Run 12 (0.16 rad/s,  16.5 m$^{-1}$, 0.189 m).}
\label{fig:A_plots}
\end{figure}

The numerical simulations and theoretical predictions are compared by measuring the viscous decay of the velocity amplitude $A_0(z)$, using the same values of $\{ k_0(z), \mu(z),\sigma(z)\}$  for the simulations and theory.  The simulation results for $A_0(z)$ as a function of the geometric length from the turning depth $(L-L_c)$  are compared to theory in Fig.\ \ref{fig:A_plots}, where $L(z) = \int_{z_0}^z \sqrt{1 + \gamma^2(z^{\prime})}dz^{\prime}$ is the distance from the wave source and $L_c \equiv L(z_c)$.  As expected, higher values of viscosity result in a more rapid decay of the velocity amplitude, as shown in Fig.\ \ref{fig:A_plots}(a).  Figure\ \ref{fig:A_plots}(b) compares cases with the same viscosity but different values of the source wavenumber, source amplitude, or wave frequency.  By comparing Run~6 to Run~9, we see that the higher wavenumber $k_0$ for Run~6 results in a more rapid viscous decay, as expected given the $k^3$ dependence in the viscous term in Eq.\ (\ref{eq:f_i}).  However, by comparing Run 12 to Run 6 we see that Run~12 decays more quickly owing to the four-fold reduction in $\omega$, even though the wavenumber for Run~12 is lower.  In all cases that we have studied, the fully-constrained theoretical predictions agree (on average) within 1.5\% of the numerical simulations, as summarized in table \ref{table:Results}.
\section{Discussion} \label{sec:Discussion}
We have presented laboratory experiments and numerical simulations of internal wave beams propagating in a fluid where the buoyancy frequency $N(z)$ varies by more than an order of magnitude and decreases exponentially as the bottom boundary is approached, as is often the case in the ocean (cf Fig.\ \ref{fig:vorticity}(a)).  We have examined wave propagation in a fluid with a turning depth $z_c$, below which $N(z) < \omega$, as occurs at many locations in the ocean \citep{king12}.  The experiments and simulations agree within a few percent with the predictions of the previously untested linear theory of \citet{kistovich98} for internal wave beam propagation in a fluid with arbitrary stratification. 

The analytical expressions describing the viscous decay of internal waves in nonuniform stratifications, which we have confirmed here, highlight the fact that beams with higher wavenumbers will decay much more rapidly owing to the $k^3$ dependence in the viscous decay term.  Critical and supercritical topography are expected to generate the strongest internal tides in the ocean, and thereby play a dominant role in models.  However, internal tides produced by such steep topography are also typically characterized by a high wavenumber, which could therefore lead to significant viscous decay.   

The energy flux below a turning depth is shown to decay exponentially with a decay constant $\alpha = k_c$, where $k_c$ is the horizontal wavenumber at the turning depth.  This decay may limit the role of internal tides beneath turning depths.  For example, the energy flux for an evanescent wave beam with a wavelength $\lambda \sim 300$~m will decay by an order of magnitude every $\sim 100$~m beneath the turning depth.  \citet{king12} found that turning depths can occur at depths greater than 1~km above the ocean bottom.  Another consequence of the existence of turning depths well above the ocean floor is the effect on internal tide generation by flow over topography in a region where internal waves are evanescent.  Internal tides generated below a turning depth will be greatly reduced in intensity by the exponential damping, which has not been considered in present models. 

The present work has examined 2D wave reflection and transmission at a turning depth for small amplitude waves at low Reynolds number and large Schmidt number.  Future work should extend to larger amplitude waves to examine the effect of nonlinearity; a recent study of internal wave reflection from a sloping bottom boundary revealed that even a small amount of nonlinearity led to large departures from the predictions of linear theory \citep{rodenborn11}.  An important but challenging extension of this work should examine the effect of a turning depth on internal wave behavior at higher Reynolds number and lower Schmidt number, where wave breaking can occur. Such a study, necessarily three-dimensional, could provide insight into the poorly understood problem of mixing in the global oceans \citep{munk98,wunsch04}.  
\section{Acknowledgments} 
We thank Bruce Rodenborn, Philip Morrison and Robert Moser for insightful discussions and support from the Office of Naval Research MURI Grant N000141110701.


\newpage
\begin{table}
  \begin{center}
\def~{\hphantom{0}}
  \begin{tabular}{lccccc}
      Run  & $\omega$~(rad/s)   &   $\nu$~($\times 10^{-6}$~m$^2$/s) & $A_0(z_0)$~(mm/s) & $k_0(z_0)$~(m$^{-1}$) & $\sigma(z_0)$~(m) \\[3pt]
      1 & 0.63 & 1 & 0.29 & 24.1 & 0.109\\
      2 & 0.63 & 2 & 0.27 & 23.8 & 0.111\\
      3 & 0.63 & 5 & 0.22 & 22.0 & 0.124\\
      4 & 0.63 & 10 & 0.18 & 20.7 & 0.135\\
      5 & 0.63 & 1 & 1.37 & 23.2 & 0.123\\
      6 & 0.63 & 5 & 1.13 & 22.3 & 0.127\\
      7 & 0.63 & 5 & 3.38 & 18.4 & 0.157\\
      8 & 0.63 & 5 & 0.12 & 14.6 & 0.176\\
      9 & 0.63 & 5 & 0.16 & 18.7 & 0.144\\
      10 & 0.63 & 5 & 0.25 & 25.5 & 0.117\\
      11 & 0.63 & 5 & 0.09 & 25.9 & 0.137\\
      12 & 0.16 & 5 & 0.33 & 16.5 & 0.189\\
       \end{tabular}
  \caption{Parameters describing the numerical simulations performed with Grid II and $\rho_{\mr{sim}}$, used to compare to the theory of  \citet{kistovich98}.  Horizontal cross-sections of the vertical velocity field near the wave source are given by $w(x,t) = A_0(z_0) \cos \left( k_0(z_0)x - \omega t \right) \exp \left[-(x-\mu(z_0))^2/2\sigma(z_0)\right]$, where $A_0$ is the velocity amplitude, $\omega$ the angular frequency, $k_0$ the wavenumber and $\sigma$ the beam width.  The kinematic viscosity $\nu$ is constant throughout the simulation domain.}\label{table:Params}
  \end{center}
\end{table}

\begin{table}
  \begin{center}
\def~{\hphantom{0}}
  \begin{tabular}{lcccc}
      Run  & $\Phi (z_c)$ ($10^{-7}$ kg/s$^3$) & $k_c$~(m$^{-1}$) & $\alpha$~(m$^{-1}$) & $\delta$ ($\%$) \\[3pt]
      1 & -3.82 & 22.2 & 22.7 & 0.48\\
      2 & -4.15 &  21.0 & 22.1 & 1.08\\
      3 & -3.63 & 18.6 & 20.3 & 0.96\\
      4 & -2.52 & 17.0 & 18.2 & 2.08\\
      5 & -107 & 22.2 & 21.8 & 2.28\\
      6 & -92.6 & 18.6 & 20.2 & 0.50\\
      7 & -1721 & 18.2 & 17.9 & 0.74\\
      8 & -1.81 & 14.6 & 15.5 & 3.42\\
      9 & -2.70 &17.0 & 17.7 & 0.86\\
      10 & -3.89 & 20.9 & 20.0 & 2.77\\
      11 & -0.91 & 23.7 & 25.2 & 2.20\\
      12 & -4.90 & 14.0 & 15.3 & 1.39\\
      \end{tabular}
  \caption{Summary of results presented in Figs.\ \ref{fig:flux_plots} and \ref{fig:A_plots}, where $\Phi(z_c)$ is the horizontally-integrated vertical flux at the turning depth, $k_c$ is the horizontal wavenumber at the turning depth, $\alpha$ is the decay constant describing the exponential decay of the vertical flux below the turning depth, and $\delta$ is the mean absolute percent difference between our measurements of $A_0(z)$ and the theoretical predictions of \citet{kistovich98}.}\label{table:Results}
  \end{center}
\end{table}

\begin{table}
  \begin{center}
\def~{\hphantom{0}}
  \begin{tabular}{lc}
    $z$ (m) & $\rho_{\mr{expt1}}$ (kg/m$^{3}$)\\
    0 & 1047.7\\
    0.0246 & 1047.3\\
    0.0819 & 1046.8\\
    0.1305 & 1045.7\\
    0.1728 & 1043.7\\
    0.2102 & 1041.6\\
    0.2438 & 1039.4\\
    0.2742 & 1037.4\\
    0.3020 & 1035.2\\
    0.3276 & 1033.1\\
    0.3513 & 1031.0\\
    0.3735 & 1028.8\\
    0.3942 & 1026.7\\
    0.4136 & 1024.6\\
    0.4320 & 1022.4\\
    0.4494 & 1020.0\\
    0.4659 & 1018.2\\
    0.4816 & 1016.0\\
    0.4965 & 1013.7\\
    0.5108 & 1011.7\\
   \end{tabular}
  \caption{Measurements of the experimental density profile $\rho_{\mr{expt1}}$, which is best fit by Eq. \ref{eq:rho_I}.}\label{table:rho_expt1}
  \end{center}
\end{table}  

\begin{table}
  \begin{center}
\def~{\hphantom{0}}
  \begin{tabular}{lc}
    $z$ (m) & $\rho_{\mr{expt2}}$ (kg/m$^{3}$)\\
    0 & 1053.7\\
    0.1885 & 1053.6\\
    0.2619 & 1051.4\\
    0.3003 & 1049.2\\
    0.3265 & 1047.0\\
    0.3464 & 1045.1\\
    0.3624 & 1042.9\\
    0.3759 & 1040.9\\
    0.3875 & 1039.0\\
    0.3976 & 1037.0\\
    0.4067 & 1035.1\\
    0.4149 & 1032.3\\
    0.4223 & 1031.0\\
    0.4291 & 1029.3\\
    0.4354 & 1027.5\\
    0.4413 & 1025.4\\
    0.4468 & 1023.5\\
    0.4520 & 1021.7\\
    0.4571 & 1020.0\\
    0.4622 & 1018.0\\
    0.4673 & 1016.5\\
    0.4724 & 1014.6\\
    0.4775 & 1012.8\\
    0.4826 & 1011.2\\
    0.4877 & 1009.3\\
    0.4928 & 1007.9\\
    0.4979 & 1006.5\\
    0.5030 & 1005.3\\
    0.5081 & 1004.0\\
       \end{tabular}
  \caption{Measurements of the experimental density profile $\rho_{\mr{expt2}}$, which is best fit by Eq. \ (\ref{eq:rho_expt2})}\label{table:rho_expt2}
  \end{center}
\end{table}  
  
\end{document}